\documentclass{PoS}

\usepackage{tikz}
\usetikzlibrary{arrows,shapes,calc}

\title{Tracks of resonances in electroweak effective Lagrangians}

\ShortTitle{Tracks of resonances in electroweak effective Lagrangians}

\author{\speaker{Ignasi Rosell}\thanks{We wish to thank the organizers for the pleasant conference. This work has been supported in part by the Spanish Government and ERDF funds from the European Commission (FPA2014-53631-C2-1-P, FPA2016-75654-C2-1-P); by the Spanish Centro de Excelencia Severo Ochoa Programme (SEV-2014-0398); by the Universidad CEU Cardenal Herrera (INDI16/10) and by La Caixa (Ph.D. grant for Spanish universities). Preprint numbers: IFIC/17-53, FTUV/17-1018.}\\
        Departamento de Matem\'aticas, F\'\i sica y Ciencias Tecnol\' ogicas, Universidad Cardenal Herrera-CEU, CEU Universities, 46115 Alfara del Patriarca, Val\`encia, Spain \\
        E-mail: \email{rosell@uchceu.es}}
        
\author{Claudius Krause, Antonio Pich and Joaqu\'\i n Santos \\
Departament de F\'\i sica Te\`orica, IFIC, Universitat de Val\`encia -- CSIC, Apt. Correus 22085, 46071 Val\`encia, Spain\\   E-mail: \email{claudius.krause@ific.uv.es, pich@ific.uv.es, joaquin.santos@ific.uv.es}}
    
    \author{Juan Jos\'e Sanz-Cillero \\ Departamento de F\'\i sica Te\'orica I, Universidad Complutense de Madrid, 28040 Madrid, Spain \\ E-mail: \email{jusanz02@ucm.es}}

\abstract{Taking into account the negative searches for New Physics at the LHC, electroweak effective theories are appropriate to deal with current energies. Tracks of new, higher scales can be studied  through next-to-leading order corrections of the electroweak effective theory. We assume a generic non-linear realization of the electroweak symmetry breaking with a singlet Higgs and a strongly-coupled UV-completion. We further consider a high-energy Lagrangian that incorporates explicitly a general set of new heavy fields. After integrating out these heavy resonances, we study the pattern of low-energy constants among the light fields, which are generated by the massive states.}

\FullConference{The European Physical Society Conference on High Energy Physics\\
		5-12 July, 2017\\
		Venice}

\begin{document}

\section{Introduction}

With all the results at hand coming from the LHC, the Standard Model (SM) describes successfullly, so far, the electroweak and strong interactions. Moreover, direct searches for New Physics (NP) states have given negative results, so there is a mass gap between SM fields and possible NP states. This mass gap justifies the use of effective field theories (EFT)~\cite{EFT} at current energies. The main idea of this work is to study the tracks of these possible higher scales through next-to-leading order corrections in the effective theory~\cite{PRD,JHEP,now}. We consider high-energy Lagrangians which incorporate these new resonances, and then we integrate out these new fields in order to estimate the low-energy couplings (LECs) in terms of resonance parameters. In this way, we can get information of the underlying theory by looking at the phenomenology at low energies, {\it i.e.}, the bottom-up approach is followed. Since direct experimental searches for these new scales are not possible yet, we can get to know these possible resonances by analyzing the pattern of the LECs of the low-energy EFT.

Depending on the nature of the electroweak symmetry breaking (EWSB), {\it i.e.}, basically depending on the way the ``Higgs'' is ``considered'', there are different possibilities for these EFTs at low energies~\cite{Claudius}:
\begin{itemize}
\item In the standard model effective theory (SM-EFT) we assume a SM Higgs at low energies. Consequently, this field forms a complex doublet representation with the Goldstone bosons of the EWSB. It usually describes weakly coupled scenarios and is organized through an expansion in canonical dimensions, in such a way that the SM Lagrangian corresponds to the leading order (LO) Lagrangian of the SM-EFT. Since Goldstone bosons in the doublet transform linearly under gauge transformations, the SM-EFT is also called "linear" EFT.
\item In the electroweak effective theory (EWET), also known as electroweak chiral Lagrangian (EWChL) or Higgs effective field theory (HEFT), one does not assume any relation between the Higgs and the Goldstones, and the Higgs is formally introduced as a scalar singlet. It usually describes strongly-coupled physics and  there is an expansion in loops or chiral dimensions~\cite{weinberg}, in such a way that the renormalizable massless (unbroken) SM Lagrangian plus the terms related to the EWSB (incorporating the Higgs and the Goldstone boson) constitute the LO Lagrangian of the EWET. In this work we make use of this more general framework, also called ``non-linear'' EFT.
\end{itemize}

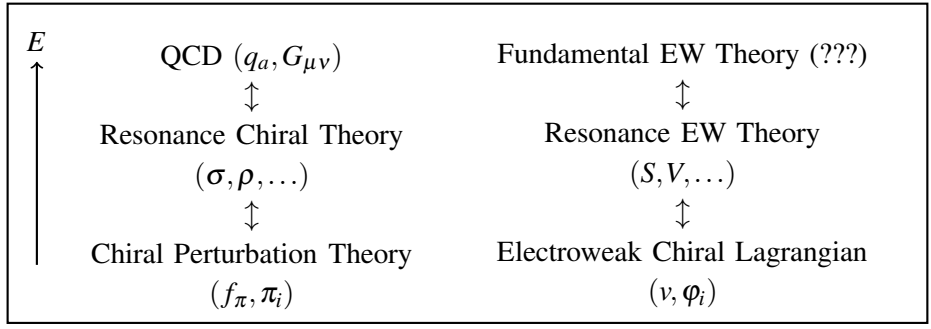
\begin{figure}[!h]
   \begin{center}
     \begin{tikzpicture}
       \draw[thick] {(0em,0em) rectangle (0.8\textwidth,11em)};
       \draw[->,thick] ($(1em,2em)$) -- ($(1em,9em)$) node [above] {$E$};
       \node [text width = 15em, anchor = north west, text centered]
(qcd) at ($(0.5em,10em)$) {QCD $(q_{a}, G_{\mu\nu})$\\ $\updownarrow $\\
Resonance Chiral Theory\\ $(\sigma, \rho, \dots)$\\ $\updownarrow
$\\Chiral Perturbation Theory\\ $(f_{\pi}, \pi_{i})$ };
       \node [text width = 15em, anchor = north east, text centered]
(ewXL) at ($(0.8\textwidth,11em)+(-0.5em,-1em)$) {Fundamental EW Theory
(???)\\ $\updownarrow $\\ Resonance EW Theory\\ $(S, V, \dots)$\\
$\updownarrow $\\Electroweak Chiral Lagrangian\\ $(v, \varphi_{i})$};
     \end{tikzpicture}
   \end{center}
   \caption{Schematic view of the energy regime of the different
(effective) theories. We assume that the electroweak sector behaves
somehow similarly to the QCD sector, at different energies respectively.}
   \label{fig:scales}
\end{figure}

As a matter of physical motivation it is interesting to compare the scales of the EWSB and the chiral symmtry breaking (ChSB) of Quantum Chromodynamics (QCD), see Figure~\ref{fig:scales}: the pion decay constant $f_\pi=0.092\,$GeV is replaced by the electroweak scale $v=(\sqrt{2}G_F)^{-1/2}=246\,$GeV. A na\"\i ve rescaling from QCD, $M_\rho=0.77\,$GeV and $M_{a1}=1.3\,$GeV, to the electroweak sector would imply vector and axial-vector resonances of $2.1\,$TeV and $3.4\,$TeV respectively. Note that the TeV scale would be interesting for the LHC phenomenology. Therefore, and following the parallelism between these QCD and electroweak frameworks, we keep track of our previous works in QCD~\cite{ChPT_RChT}, where we determined some LECs of Chiral Perturbation Theory (ChPT)~\cite{weinberg,ChPT}, the very low-energy EFT of QCD, in terms of resonance parameters by using the Resonance Chiral Theory~\cite{RChT}.

A NLO analysis~\cite{PRL} of the oblique electroweak observables $S$ and $T$~\cite{Peskin} shows that this kind of strongly-coupled scenarios are phenomenologically allowed once the resonances live at the TeV scale and the $WW$ coupling of the Higgs is close to the SM value.

These proceedings are organized in the following way. In Section 2 we show the effective field theories at low energies (with only the SM particles) and at high energies (SM particles and resonances). In Section 3 we explain how to integrate out the resonances in order to determine the LECs of the EWET in terms of resonance parameters. 
We stress the improvements with respect to our previous work of Ref.~\cite{JHEP}: we also consider gluons, color octets and fermionic resonances. 

\section{The framework}

\subsection{The low-energy effective theory}

The EWET is built by considering the most general Lagrangian containing the SM fields ($W^\pm$ and $Z$ gauge bosons, gluons, fermions, electroweak Goldstones and the Higgs $h$), satisfying the SM symmetries and its pattern of EWSB: $G\equiv SU(2)_L\otimes SU(2)_R\rightarrow H\equiv SU(2)_{L+R}$. We follow the notation of Refs.~\cite{PRD,JHEP,now}:
\begin{itemize}
\item The Goldstone fields are parametrized through the $G/H$ coset representative~\cite{RChT} $u(\varphi) =\exp{(\frac{i}{2}\,\vec\sigma\vec\varphi/v)}$, transforming under chiral symmetry $g\equiv (g_L^{\phantom{\dagger}},g_R^{\phantom{\dagger}})\in G$, $u \to g_L^{\phantom{\dagger}} u \,g_h^\dagger= g_h^{\phantom{\dagger}} u \,g^\dagger_R$, with $g_h^{\phantom{\dagger}}\equiv g_h^{\phantom{\dagger}}\in H$. For convenience, we consider $U=u^2\to g_L^{\phantom{\dagger}} U g^\dagger_R$ and $u_\mu = i\, u\, (D_\mu U)^\dagger u = u_\mu^\dagger\to g_h^{\phantom{\dagger}} u_\mu g_h^{\dagger}$. 
\item The covariant derivative $D_\mu U = \partial_\mu U - i \hat W_\mu U + i U \hat B_\mu$ couples the Goldstones to external $SU(2)_{L,R}$ gauge sources, making the Lagrangian formally invariant under local $G$ transformations. The identification with the SM gauge fields, $\hat W_\mu = -\frac{g}{2}\,\vec\sigma\vec W_\mu$ and $\hat B_\mu = -\frac{g'}{2}\,\sigma_3 B_\mu$, breaks explicitly the $SU(2)_R$ symmetry while preserving the $SU(2)_L\otimes U(1)_Y$ SM symmetry. 
\item The left and right field-strength tensors have been re-written in terms of $f_\pm^{\mu\nu}\equiv u^\dagger \hat W^{\mu\nu} u \pm u\, \hat B^{\mu\nu} u^\dagger$, which transform as triplets under $G$: $f_\pm^{\mu\nu} \to g_h^{\phantom{\dagger}} f_\pm^{\mu\nu} g_h^{\dagger}$. 
\item With quarks the symmetry group must be extended to $G\equiv SU(3)_C\otimes SU(2)_L\otimes SU(2)_R \otimes U(1)_X$. The quarks transform under $G$ like $\psi_L  \rightarrow  g_C g_X  g_L  \psi_L $ and $\psi_R \rightarrow g_C g_X g_R  \psi_R$ with $g_X \in U(1)_X$ and $g_C \in SU(3)_C$. 
In the Lagrangian we introduce the fermions by considering the bilinears $J_\Gamma$ 
(color singlet) and $J_\Gamma^8 \equiv \big( J^{8,a}_\Gamma \big) T^a$ 
(color octet), with $\Gamma = 1$, $i\gamma_5$, $\gamma_\mu$, $\gamma_\mu \gamma_5$ and $\sigma_{\mu \nu}$ for the scalar (S), pseudoscalar (P), vector (V), axial-vector (A) and tensor (T) cases.   
\item The covariant derivatives of the fermions introduce the auxiliary $U(1)_X$ field $\hat{X}_\mu$, associated with $\mathrm{B}-\mathrm{L}$ and responsible of the electric charge difference $|Q_u| \neq |Q_d|$ and $|Q_e|\neq |Q_\nu|$. The related strength tensor $\hat{X}_{\mu\nu}$ is a singlet under $G$ and one recovers the SM gauge interactions with the identification $\hat{X}_\mu =- g' B_\mu$. 
\item The covariant derivatives of the quarks introduce also the gluon fields $\hat{G}^\mu=g_S G^a_\mu T^a$, with $g_S$ the strong coupling and $T^a$ the $SU(3)$ generator. The field strength tensor is given by $G^a_{\mu\nu}$ and $\hat{G}_{\mu\nu}=g_S G^a_{\mu\nu} T^a$.
\item We use $\mathcal{T}$ to introduce an explicit breaking of custodial symmetry, $\mathcal{T}= u \mathcal{T}_R u^\dagger \rightarrow\quad g_h^{\phantom{\dagger}} \mathcal{T} g_h^\dagger$, being $\mathcal{T}_R$ the right-handed spurion $\mathcal{T}_R\rightarrow g_R^{\phantom{\dagger}} \mathcal{T}_R g_R^\dagger$ and making the identification $\mathcal{T}_R  = -g'\frac{\sigma_3}{2}$ in order to recover the SM. This tensor appears, for instance, in the Yukawa mass term and is reponsible for the up-down mass splitting.
\end{itemize}

The effective Lagrangian is organized as a low-energy expansion in powers of momenta, 
\begin{equation}
\mathcal{L}_{\mathrm{EWET}} \,=\, \sum_{\hat d\ge 2}\, \mathcal{L}_{\mathrm{EWET}}^{(\hat d)}\,, \label{EWET-Lagrangian0}
\end{equation}
where the operators are not simply ordered according to their canonical dimensions and one must use instead the dimension $\hat d$ which reflects their infrared behaviour at low momenta~\cite{weinberg}. Quantum loops are renormalized order by order in this low-energy expansion. The power-counting rules can be summarized as: $h/v  \sim \mathcal{O}\left(p^0\right)$; $u_\mu,\, \partial_\mu$ and $\mathcal{T} \sim \mathcal{O}\left(p\right)$; $ f_{\pm\, \mu\nu},\, \hat{X}_{\mu\nu},\,\hat{G}_{\mu\nu},$ $J_{S,P}$, $J_{V,A}^{\mu}$, $J_T^{\mu\nu}$ and $J^{8\,\mu\nu}_T  \sim \mathcal{O}\left(p^2\right)$. It is interesting to spotlight two features related to this power counting:
\begin{enumerate}
\item Assuming that the SM fermions couple weakly to the strong sector we assign an $\mathcal{O}(p^2)$ to fermion bilinears. Note that a na\"\i ve chiral analysis would have assigned an $\mathcal{O}(p)$.
\item Considering its phenomenological suppression and assuming no strong breaking of the custodial symmetry, and contrary to the pioneering papers studying the Higgsless EWET~\cite{Longhitano}, we assign an $\mathcal{O}(p)$ to the explicit breaking of this symmetry.
\end{enumerate}

As it has been pointed out previously the LO Lagrangian corresponds to the renormalizable massless (unbroken) SM Lagrangian plus the terms related to the EWSB (incorporating the Higgs and the Goldstone boson). The NLO Lagrangian~\cite{PRD,JHEP,now,Longhitano,Cata}, $\mathcal{O}(p^4)$, can be split in different pieces,
\begin{equation}
\mathcal{L}_{\mathrm{EWET}}^{(4)} \, = \, \sum_{i=1}^{12} \mathcal{F}_i \; \mathcal{O}_i  + \sum_{i=1}^3 \widetilde{\mathcal{F}}_i \; \widetilde{\mathcal{O}}_i + \sum_{i=1}^{8}  \mathcal{F}_i^{\psi^2}\; \mathcal{O}_i^{\psi^2}  + \sum_{i=1}^{3}   \widetilde{\mathcal{F}}_i^{\psi^2}\; \widetilde{\mathcal{O}}_i^{\psi^2} + \sum_{i=1}^{10}\mathcal{F}_i^{\psi^4}\; \mathcal{O}_i^{\psi^4} + \sum_{i=1}^{2} \widetilde{\mathcal{F}}_i^{\psi^4}\; \widetilde{\mathcal{O}}_i^{\psi^4} \, ,
\label{EWET-Lagrangian}
\end{equation}
where the operators have been separated considering their $P$ nature (without or with tilde for $P$-even and $P$-odd operators) and the presence of fermions. In Ref.~\cite{now} we show all the operators. Note that the different LECs can be multiplied by an arbitray polinomial of $h$~\cite{Grinstein:2007iv}, since this does not increase the counting. 

\subsection{The high-energy theory}

At higher energies we consider also resonance fields: scalar ($S$), pseudoscalar ($P$), vector ($V$) and axial-vector ($A$) bosonic resonances and fermionic resonances. 
\begin{itemize} 
\item In the case of bosonic resonances we consider generic massive states, transforming under $G$ as $SU(2)$ triplets ($R_3^m= \sigma_i R_{3,i}^m/\sqrt{2}$) or singlets ($R_{1}^m$) and as $SU(3)$ octets ($R_n^8= T^a R^{8,a}_n$) or singlets ($R_n^1$), so that $R^1_1 \rightarrow R_1^1$, $R^1_3 \rightarrow g_h^{\phantom{\dagger}} R^1_3 g_h^\dagger$, $R^8_1 \rightarrow g_C^{\phantom{\dagger}} R^8_1 g_C^\dagger$ and $R^8_3 \rightarrow g_C^{\phantom{\dagger}} g_h^{\phantom{\dagger}} R^8_3 g_h^\dagger g_C^\dagger$.  
\item In the case of fermionic resonances, we consider generic electroweak-doublet massive states $\Psi$, transforming under $G$ as a top-bottom quark doublet, with the same quantum numbers, $\Psi \rightarrow g_C\, g_X\, g_h \,\Psi$. Note that with other representations it is not possible to construct invariant operators with a single fermionic resonance and the low-energy fields. Taking into account, as we have pointed out before, that the SM fermion bilinears are $\mathcal{O}(p^2)$ because of their weak coupling, we assume that the fermion bilinears constructed between a light fermion and a heavy one are $\mathcal{O} (p)$. 
\end{itemize}

We can split the Lagrangian in terms which contain resonances explicitly, $\mathcal{L}_{\mathrm{R}}$, and terms which do not contain resonances, $\mathcal{L}_{\mathrm{non-R}}$,
\begin{equation}
\mathcal{L}_{\mathrm{RT}} \,=\, \mathcal{L}_{\mathrm{ R}} + \mathcal{L}_{\mathrm{non-R}}\, .
\label{RT-Lagrangian}
\end{equation}
Note that in $ \mathcal{L}_{\mathrm{ R}}$ we only consider terms with a single resonance, since terms with more resonances do not contribute to $\mathcal{O}(p^4)$ LECs of the EWET. The second piece on the right-hand side of (\ref{RT-Lagrangian}) is formally identical to the EWET Lagrangian of (\ref{EWET-Lagrangian0}), but with different couplings, because it describes the interactions of a different EFT, valid at the resonance mass scale. In Ref.~\cite{now} the list of operators of $\mathcal{L}_{\mathrm{R}}$ is shown: six scalar operators, five pseudoscalar operators, nineteen vector operators, nineteen axial-vector operators and six fermionic operators.

For the massive spin-1 fields there is freedom in the resonance formalism selection: we can use either the four-vector Proca formalism $\hat{R}^\mu$ or the rank-2 antisymmetric formalism $R^{\mu \nu}$. By using a change of variables in the context of a path integral formulation, and once a good short-distance behavior is required, we demonstrated the equivalence of both formalisms in Ref.~\cite{JHEP}: they lead to the same predictions for the LECs of the EWET at NLO. Although both descriptions are fully equivalent, they involve a different $\mathcal{L}_{\mathrm{R}}$ Lagrangian. Depending on the particular phenomenological application, one formalism can be more efficient than the other, in the sense that it does not receive contributions from local operators of $\mathcal{L}_{\mathrm{non-R}}$ (operators without resonances in the Resonance Effective Theory). 


\section{Determining the LECs}

As it has been explained previously, the main aim of this work is to estimate the LECs of the EWET in terms of resonance parameters. In order to do that, we need to integrate out the resonance fields and it is very convenient to consider high-energy constraints of the assumed underlying theory, so that it is well-behaved, to reduce the number of resonance parameters. The idea is that once we have the expressions of the LECs in terms of the resonance parameters, and assuming a high enough precision in the low-energy experiments, the phenomenology can give glimpses of the high-energy physics, {\it i.e.}, the New Physics. This is a typical bottom-up procedure because the high-energy effective theory is analyzed by using the NLO of the effective theory at low energies. 

It is convenient to stress that whereas resonance contributions to the LECs coming from bosonic resonances are suppressed by powers of $1/M_R^2$, the contributions of fermionic resonances are suppressed by powers of $1/M_\Psi$. Although a na\"\i ve counting would suggest that these last contributions are more important than the first ones this is not the case, since fermion currents are assumed to be weakly coupled, what introduces an additional vertex suppression and finally both contributions to the LECs share the same order.

As a first approach we studied in Ref.~\cite{PRD} the prediction of $P$-even purely bosonic $\mathcal{O}(p^4)$ LECs from $P$-even bosonic colourless resonance exchanges (without considering operators with $\mathcal{T}$ or $\hat{X}_{\mu\nu}$), once some short-distance constraints were taken into consideration, what allowed us to get predictions in terms of only a few resonance parameters. In Ref.~\cite{JHEP} we expanded our analysis to include bosonic and fermionic $\mathcal{O}(p^4)$ LECs from bosonic colourless resonance exchanges. Finally, we have considered now gluons, color octets and fermionic resonances~\cite{now}. In the future we plan to include more than one generation of quarks and/or leptons.


\begin{thebibliography}{99}

\bibitem{EFT}
  H.~Georgi,
  Ann.\ Rev.\ Nucl.\ Part.\ Sci.\  {\bf 43} (1993) 209;
  A.~Pich, Proc. Les Houches Summer School of Theoretical
Physics --Probing the Standard Model of Particle Interactions-- (Les Houches, France, 1997), eds. R. Gupta {\it et al.} (Elsevier Science B.V., Amsterdam, 1999), Vol. II, p.~949
  [hep-ph/9806303].

\bibitem{PRD}
  A.~Pich, I.~Rosell, J.~Santos and J.~J.~Sanz-Cillero,
  Phys.\ Rev.\ D {\bf 93} (2016) no.5,  055041
  [arXiv:1510.03114 [hep-ph]].

\bibitem{JHEP}
  A.~Pich, I.~Rosell, J.~Santos and J.~J.~Sanz-Cillero,
  JHEP {\bf 1704} (2017) 012
  [arXiv:1609.06659 [hep-ph]];
  EPJ Web Conf.\  {\bf 137} (2017) 10006
  [arXiv:1611.10295 [hep-ph]].

\bibitem{now}
  C.~Krause, A.~Pich, I.~Rosell, J.~Santos and J.~J.~Sanz-Cillero, work in progress.
  
  \bibitem{Claudius}
  G.~Buchalla, O.~Cata and C.~Krause,
  Nucl.\ Phys.\ B {\bf 894} (2015) 602
  [arXiv:1412.6356 [hep-ph]];
  G.~Buchalla, O.~Cata, A.~Celis and C.~Krause,
  Nucl.\ Phys.\ B {\bf 917} (2017) 209
  [arXiv:1608.03564 [hep-ph]].

\bibitem{weinberg}
  S.~Weinberg,
  Physica A {\bf 96} (1979) 327.

\bibitem{ChPT_RChT}
  I.~Rosell, J.~J.~Sanz-Cillero and A.~Pich,
  JHEP {\bf 0408} (2004) 042
  [hep-ph/0407240];
  JHEP {\bf 0701} (2007) 039
  [hep-ph/0610290];
  JHEP {\bf 0807} (2008) 014
  [arXiv:0803.1567 [hep-ph]];
  JHEP {\bf 1102} (2011) 109
  [arXiv:1011.5771 [hep-ph]].

\bibitem{ChPT}
  J.~Gasser and H.~Leutwyler,
  Annals Phys.\  {\bf 158} (1984) 142;
  Nucl.\ Phys.\ B {\bf 250} (1985) 465,
%
  517,
%
  539; 
  J.~Bijnens, G.~Colangelo and G.~Ecker,
  JHEP {\bf 9902} (1999) 020
  [hep-ph/9902437];
  Annals Phys.\  {\bf 280} (2000) 100
  [hep-ph/9907333].

\bibitem{RChT}
  G.~Ecker, J.~Gasser, A.~Pich and E.~de Rafael,
  Nucl.\ Phys.\ B {\bf 321} (1989) 311; 
  G.~Ecker, J.~Gasser, H.~Leutwyler, A.~Pich and E.~de Rafael,
  Phys.\ Lett.\ B {\bf 223} (1989) 425; 
  V.~Cirigliano, G.~Ecker, M.~Eidemuller, R.~Kaiser, A.~Pich and J.~Portol\'es,
  Nucl.\ Phys.\ B {\bf 753} (2006) 139
  [hep-ph/0603205].

\bibitem{PRL}
  A.~Pich, I.~Rosell and J.~J.~Sanz-Cillero,
  Phys.\ Rev.\ Lett.\  {\bf 110} (2013) 181801
  [arXiv:1212.6769];
  JHEP {\bf 1401} (2014) 157
  [arXiv:1310.3121 [hep-ph]].

\bibitem{Peskin}
  M. E. Peskin and T. Takeuchi, Phys.\ Rev.\ Lett.\  {\bf 65} (1990) 964;
  Phys.\ Rev.\ D {\bf 46} (1992) 381.

\bibitem{Longhitano}
  A.~C.~Longhitano,
  Phys.\ Rev.\ D {\bf 22} (1980) 1166;
%
  Nucl.\ Phys.\ B {\bf 188} (1981) 118.

\bibitem{Cata}
  G.~Buchalla, O.~Cat\`a and C.~Krause,
  Nucl.\ Phys.\ B {\bf 880} (2014) 552
  [arXiv:1307.5017 [hep-ph]]; 
  G.~Buchalla and O.~Cat\`a,
  JHEP {\bf 1207} (2012) 101
  [arXiv:1203.6510 [hep-ph]].

\bibitem{Grinstein:2007iv}
  B.~Grinstein and M.~Trott,
  Phys.\ Rev.\ D {\bf 76} (2007) 073002
  [arXiv:0704.1505 [hep-ph]].


\end{thebibliography}
\end{document}